\documentclass[letterpaper, 10 pt, conference]{ieeeconf}  
\IEEEoverridecommandlockouts                              
\usepackage{verbatim}
\usepackage{amsfonts,amsmath,mathrsfs,amssymb,amsbsy}
\usepackage[final]{graphicx}
\usepackage{times,cite}
\usepackage{caption}
\usepackage{subcaption}
\usepackage{url}
\usepackage{balance}

\long\def\symbolfootnote[#1]#2{\begingroup%
\def\thefootnote{\fnsymbol{footnote}}\footnote[#1]{#2}\endgroup}

\newcommand{\Prob}{\mathsf{P}}
\newcommand{\Expect}{\mathsf{E}}

\newcommand{\tauc}{\tau_{\scriptscriptstyle \text{C}}}
\newcommand{\taug}{\tau_{\scriptscriptstyle \text{G}}}

\title{\LARGE \bf Sequential Detection of Regime Changes in Neural Data
}

\author{Taposh Banerjee, Stephen Allsop, Kay M. Tye, Demba Ba and Vahid Tarokh
\thanks{Banerjee is with the University of Texas at San Antonio (ECE), Allsop is with the Harvard Medical School, Tye is with the MIT Dept. of Brain and Cognitive Sciences, 
Ba is with Harvard University (SEAS), and Tarokh is with Duke University (ECE). Corresponding author email: taposh.banerjee@utsa.edu.}
\thanks{Banerjee, Ba and Tarokh acknowledges support of the Army Research Office under Contract Number W911NF-16-1-0368.  This is part of the collaboration between US DOD, UK MOD and UK Engineering and Physical Research Council (EPSRC) under the Multidisciplinary University Research Initiative. Tye acknowledges support of the Pioneer Award DP1-AT009925 (NCCIH). Allsop acknowledges support of the Jeffrey and Nancy Halis Fellowship, the Henry E. Singleton Fund, and an NLM training grant. }
}

\begin{document}

\maketitle
\thispagestyle{empty}
\pagestyle{empty}

\begin{abstract}
The problem of detecting changes in firing patterns in neural data is studied.  
The problem is formulated as a quickest change 
detection problem. Important algorithms from the literature are reviewed. A new algorithmic technique is discussed to detect
deviations from learned baseline behavior. The algorithms studied can be applied to both spike and local field potential data. 
The algorithms are applied to mice spike data to verify the presence of behavioral learning. 
\end{abstract}

\section{INTRODUCTION}
A major research direction in the area of brain-computer interfaces (BCIs) is the decoding of brain signals \cite{rao2013brain}. 
Most of the research in this area focus on classification of brain signals based on spike data or local field potential (LFP) data \cite{markowitz2011optimizing}, \cite{bane-ssp-2018}, \cite{bane-icassp-2018}. The classification algorithms studied in the literature are based on the idea that the firing pattern of neurons will be different under different classes. 
The different firing patterns affect the number and positions of spikes in the spike data, and may also affect the frequency spectrum of the LFP data. The latter fact is used 
to train classifiers based on the Fourier or wavelet coefficients, or based on the power spectrum of the LFP data. 

In this paper, we study another important aspect of brain signal processing that is of detecting changes in neural firing patterns. 
For future BCIs, it is envisioned that the BCIs will be capable of decoding brain signals and controlling external outputs, e.g., a robotic arm. In these applications, while 
is important to decode what the brain is trying to do in a given task, it is also important to learn the transition boundaries between different types of tasks. Thus, algorithms 
are needed that can observe spike or LFP data in real-time and detect changes in regimes in neural data. 
Such change detection algorithms can also be used for classification purposes. For example, a change in firing pattern compared to a baseline pattern may be used to test a hypothesis, for example, to verify a change in behavior or to test if an animal learned to associate an activity to a cue, etc. 

In this paper, we study algorithms that can be used to detect changes in statistical behavior of spike and LFP data. In Statistics, such change detection algorithms are
developed in the framework of quickest change detection \cite{poor-hadj-qcd-book-2009}, \cite{tart-niki-bass-2014}, \cite{veer-bane-elsevierbook-2013}. 
In Section~\ref{sec:QCDreview}, we review some fundamental algorithms from the literature. As will be discussed, to effectively use the algorithms, 
we need knowledge of both the pre- and post-change distributions of the data. In Section~\ref{sec:QCDDeviation}, we propose change detection algorithms that 
can be used to detect deviations from learned baseline behavior, and hence can be used without knowledge of the post-change distribution. In Section~\ref{sec:MiceData}, we apply the algorithms to spike data to verify behavioral learning in mice.

\section{Review of Quickest Change Detection}\label{sec:QCDreview}

\begin{figure}
\centering
\includegraphics[scale=0.4]{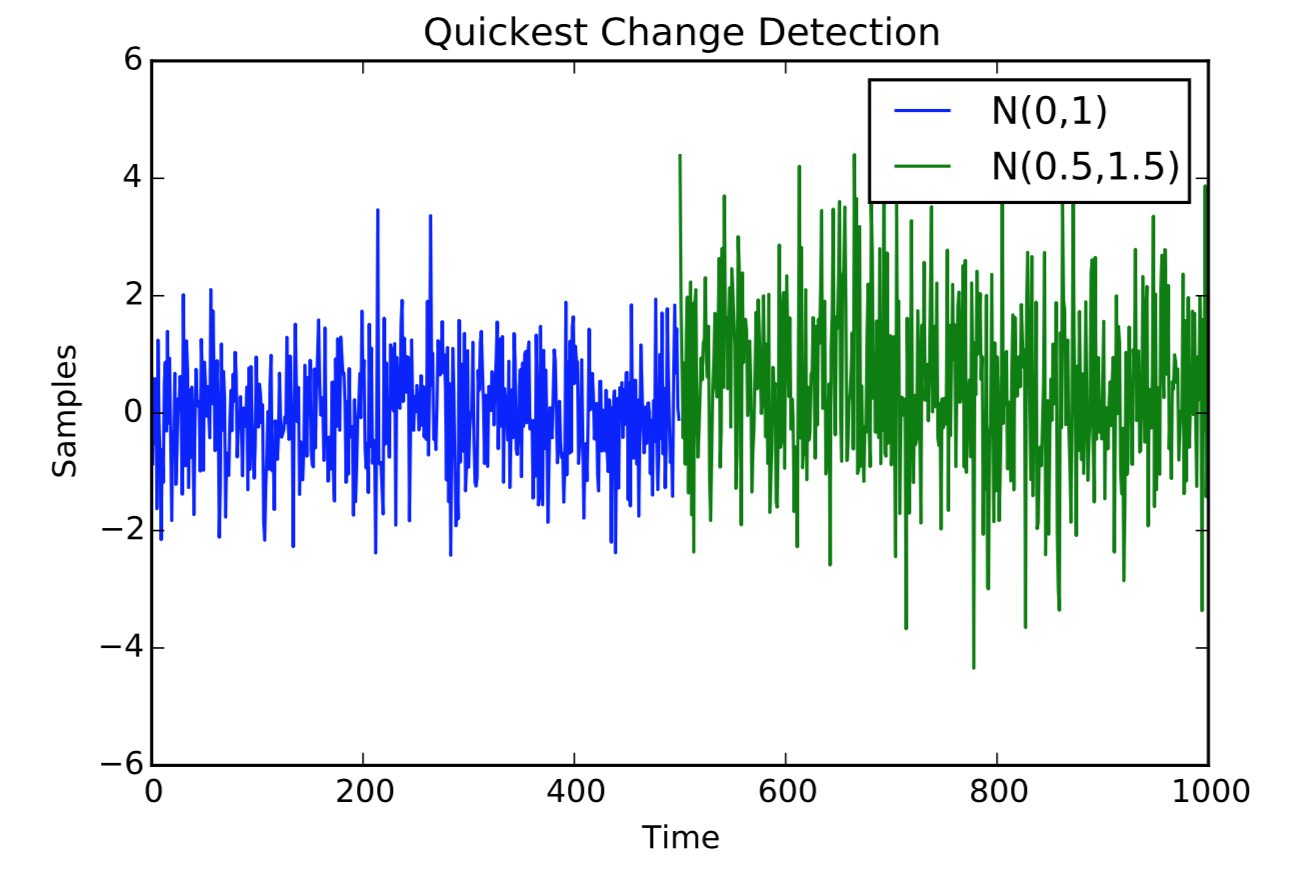}
\caption{A change in the mean and variance of a sequence of independent Gaussian random variables.}
\label{fig:QCD}
\end{figure}

We model the spike data or the LFP data as a stochastic process $\{X_n\}$ with probability law in a parametric family $\Prob_\theta$. Here, $\theta$ could be 
infinite dimensional making the probability law nonparametric. We assume that in the nominal regime the law of 
the process is $\Prob_{\theta_0}$. We assume that at some point in time $\gamma$ called the change point in the following, the law 
of the process changes to $\Prob_{\theta_1}$. See Fig.~\ref{fig:QCD}. The objective is to observe this process in real-time and detect this change in the law as quickly as possible. 
The algorithm to detect this change is expressed in terms of a stopping time $\tau$: an integer-valued random variable such that the event $\{\tau \leq n\}$ is only a function 
of the first $n$ observations $(X_1, \cdots, X_n)$. The variable $\tau$ has to be selected so as to minimize a version of the delay $\tau - \gamma$ subject to a constraint on the event of false alarm $\{\tau < \gamma\}$. The random variables can be dependent and the parameters $\theta_0$ and $\theta_1$ may not be known. In the subsections below we discuss change detection algorithms that are effective under various modeling assumptions. 

\subsection{IID Data with Known $\theta_0$ and $\theta_1$}
We assume that the random variables are i.i.d. 
with probability density function (p.d.f.) $f_0$ (under law $\Prob_{\theta_0}$ observations are i.i.d. with density $f_0$).
At time $\gamma$ the density of the random variables changes to $f_1 \neq f_0$ 
(under law $\Prob_{\theta_1}$ observations are i.i.d. with density $f_1$). 
Thus, the variables $\{X_n\}$ are independent conditioned on the change point $\gamma$. 
There are many popular formulations of the QCD problem, the most popular once are the formulations of Lorden \cite{lord-amstat-1971} and Pollak \cite{poll-astat-1985}. We do not discuss the problem formulations here. But, an algorithm that is optimal in some well-defined sense 
with respect to both these formulations is the Cumulative Sum (CUSUM) algorithm. The CUSUM algorithm was proposed by Page 
in \cite{page-biometrica-1954}. It is described as follows. We compute a sequence of statistics $\{W_n\}$ using the log likelihood ratio 
of the observations: 
\begin{equation}\label{eq:CUSUM_stat_ML}
W_n = \max_{1 \leq k \leq n+1} \sum_{i=k}^n \log \frac{f_1(X_i)}{f_0(X_i)},
\end{equation}
and a change is declared, i.e., 
and an alarm is raised, the first time the statistic is above a threshold $A$: 
\begin{equation}\label{eq:CUSUM_stop}
\tauc = \min \{n \geq 1: W_n > A\}.
\end{equation}
The statistic $W_n$ is a maximum likelihood statistic: the term $\sum_{i=k}^n \log f_1(X_i)/f_0(X_i)$ is the log likelihood 
ratio of the observations given $\gamma=k$. The statistic $W_n$ is the maximum of this conditioned log likelihood ratio over
all possible change points ${1 \leq k \leq n}$ before $n$ and $k=n+1$. The latter represents no change for which the log likelihood ratio is zero keeping the statistic positive.    
The statistic $W_n$ can be computed recursively as follows: $W_0=0$, and 
\begin{equation}\label{eq:CUSUM_stat_Recur}
W_n = \left( W_{n-1} + \log \frac{f_1(X_n)}{f_0(X_n)} \right)^+,
\end{equation}
where $(x)^+ := \max \{x, 0\}$. 
It is possible that a change never occurs ($\gamma=\infty$). In that case, $\tauc$ is the time to false alarm and can be controlled by the threshold $A$. 
Thus, the threshold $A$ provides a trade-off between delay and false alarm because a larger value of threshold also leads to a larger delay 
when the change actually occurs. We refer the readers to \cite{veer-bane-elsevierbook-2013} and \cite{lai-ieeetit-1998} for delay and false alarm analysis of the CUSUM algorithm. 

To understand why the CUSUM algorithm works, define the notion of Kullback-Leibler divergence between probability densities $f$ and $g$:
\[
D(f \;\| \; g) = \int f(x) \log \frac{f(x)}{g(x)} dx.
\]   
It is well known that \cite{veer-bane-elsevierbook-2013}
\[
D(f \;\| \; g) \geq 0, \mbox{ with equality iff } f=g.
\]
At each time step, the log likelihood ratio of the observations $\log f_1(X_n)/f_0(X_n)$ 
is added to the statistic $W_n$. If there is no change or anomaly, then the observations
have density $f_0$, and the average value of the log likelihood ratio under $f_0$ is $-D(f_0 \| f_1) < 0$. 
After the change, the observations have density $f_1$ and the mean of log likelihood ratio is $D(f_1 \| f_0) > 0$. 
Thus, before the change, the statistic $W_n$ has a tendency to go to $-\infty$ (but is stopped at $0$ by the $(\cdot)^+$ operation). 
After the change, the statistic $W_n$ has a tendency to grow to $\infty$, this is detected by using a suitable large threshold of $A$.   

If the spike data is modeled as a Poisson or a Bernoulli process with known pre- and post-change parameters, then we can use the CUSUM algorithm to detect 
a change in the firing rate or pattern by detecting a change in Poisson or Bernoulli parameters.

\subsection{IID Data with $\theta_0$ and $\theta_1$ Unknown but Finite Dimensional}
The CUSUM algorithm can be applied only when both the pre- and post-change densities $f_0$ and $f_1$ are precisely known.
If $f_0$ and $f_1$ are not known to us beforehand and has to be estimated based on some training data, then the CUSUM algorithm 
is no longer optimal. In fact, the algorithm may even fail to detect changes accurately due to the error in estimating 
the densities $f_0$ and/or $f_1$. In such a situation we can use the \textit{Generalized CUSUM} (GCUSUM)
algorithm based on the Generalized Likelihood Ratio (GLR) approach. 
In the GLR approach, roughly speaking, we replace the unknown by its Maximum Likelihood (ML) estimate. The CUSUM algorithm 
is also an example of a GLR test where the GLR part is the $\max$ operation over the unknown change point. 
In the GCUSUM algorithm, in addition to a $\max$ over change point, we also replace the unknown parameters 
$\theta_0$ and $\theta_1$ by their ML estimates \cite{lord-amstat-1971}, \cite{lai-jrss-1995}, \cite{lai-ieeetit-1998}. 

We assume that the densities $f_0$ and $f_1$ belong to the parameteric family $\{f_\theta\}, \theta\in \Theta$. 
Also, let $f_0 = f_{\theta_0}$
and $f_1 = f_{\theta_1}$. We do not know the values of $\theta_0$ and $\theta_1$ but know that $\theta_0 \in \Theta_0$ and 
$\theta_1 \in \Theta_1$; $\Theta_i \subset \Theta$, $i=1,2$, $\Theta_0 \cap \Theta_1 = \emptyset$. 
Then, the GCUSUM test statistic based on observations $\{X_1, \cdots, X_n\}$ when the change occurs at $\gamma=k$ is given by 
\begin{equation}\label{eq:CUSUM_doub_GLR_k}
\begin{split}
G_n(k) =& \max_{\theta_0 \in \Theta_0} \sum_{i=1}^{k-1} \log f_{\theta_0}(X_i) + \max_{\theta_1 \in \Theta_1} \sum_{i=k}^{n} \log f_{\theta_1}(X_i) \\
& - \max_{\theta_0 \in \Theta_0} \sum_{i=1}^{n} \log f_{\theta_0}(X_i),
\end{split}
\end{equation}
and the GLR statistic at time $n$ is defined as
\begin{equation}\label{eq:CUSUM__doub_GLR}
G_n = \max_{1 \leq k \leq n } G_n(k).
\end{equation}
A change is declared at the stopping time
\begin{equation}\label{eq:CUSUM_doub_GLR_stop}
\taug = \min\{n \geq 1: G_n > A\}.
\end{equation}

The above GCUSUM test is discussed in \cite{lai-jrss-1995}. Although the test performs well in practice, 
it is known to be optimal or asymptotically optimal only if some additional assumptions are made about the family of densities. One specific case is when the densities belong to an exponential family, and the pre-change distribution is known. Such an analysis was carried out by Lorden in \cite{lord-amstat-1971}. We refer the readers to 
\cite{bane-hero-TSP-2018} for more details on the Lorden's test and its detailed analysis under misspecfication of the pre-change distribution. 
There are also other approaches and algorithms using which one can detect changes under model uncertainty. We refer the readers to \cite{veer-bane-elsevierbook-2013} for a review. 

If the spike data is modeled as a Poisson or a Bernoulli process with unknown pre- and post-change parameters, then we can use the GCUSUM algorthm to detect 
a change in the firing rate or pattern by detecting a change in Poisson or Bernoulli parameters. 

\subsection{Algorithm for Dependent Data}
Both the CUSUM algorithm and the GCUSUM algorithm are designed to work with i.i.d. data. 
In general, the data sequence need not be i.i.d., and we need more general algorithms to detect changes. 
Algorithms for non-i.i.d. data can be obtained by replacing the product densities in the definition of the CUSUM or GCUSUM algorithms
by joint densities. 
 
Let $X_1^n$ denote the vector $(X_1, \cdots, X_n)$. Also, let $f^{(k)}(x_1^n)$ denote the joint density of $X_1^n$ given 
that change occured at time $\gamma=k$. Then the CUSUM statistic for non-i.i.d. data is given by (compare with \eqref{eq:CUSUM_stat_ML})
\begin{equation}\label{eq:CUSUM_stat_noniid}
W_n = \max_{1 \leq k \leq n+1} \log \frac{f^{(k)}(X_1^n)}{f^{(\infty)}(X_1^n)}.
\end{equation}
Let the joint density of $X_1^n$ under law $\Prob_{\theta}$ be of the form
\[
f_\theta(x_1^n) =  \prod_{i=1}^n f(x_i | x_1^{i-1}; \theta).
\]
Then, one way to write the CUSUM statistic for the non-i.i.d. data above is 
\begin{equation}\label{eq:CUSUM_stat_noniid_withden}
W_n = \max_{1 \leq k \leq n+1} \sum_{i=k}^n \log \frac{f(X_i | X_1^{i-1}; \theta_1)}{f(X_i | X_1^{i-1} ; \theta_0)}.
\end{equation}
Another way is to assume independent pre- and post-change data
\begin{equation}\label{eq:CUSUM_stat_noniid_withden_ind}
W_n = \max_{1 \leq k \leq n+1} \sum_{i=k}^n \log \frac{f(X_i | X_k^{i-1}; \theta_1)}{f(X_i | X_k^{i-1} ; \theta_0)}.
\end{equation}

If the spike data is modeled as a hidden Markov model or a state-space process, then we can use the modified CUSUM algorithm for dependent data to detect 
a change in the firing pattern. Also, if the LFP data is modeled as a ARMA time-series model, then also we can use algorithms in \eqref{eq:CUSUM_stat_noniid_withden}
or \eqref{eq:CUSUM_stat_noniid_withden_ind} to detect the change.  

\section{Detecting Deviations from a Known Baseline Behavior}\label{sec:QCDDeviation}
In order to employ the CUSUM and the GCUSUM algorithm, we need prior information on the post-change distribution: we 
need to know the exact post-change distribution for the CUSUM algorithm and need to know the post-change parametric family for the GCUSUM algorithm.  
In many applications, we may not always know the statistical characteristics of the data in the anomalous regime or have too few samples from the anomalous regime to learn 
the post-change distribution. In this section, we discuss an algorithms that we can employ in such a scenario. 

Suppose we have a summary statistics $h: \mathbb{R}^d \to \mathbb{R}$ of the data such that
\begin{equation}
\Expect[h(X_{n-d}, \cdots, X_n)] = \mu_0, \quad \forall n, 
\end{equation}
in the nominal regime, and we expect this mean to increase in the post-change regime. Then we can use
\begin{equation}
\begin{split}
W_n &= (W_{n-1} + h(X_{n-d}, \cdots, X_n) - \mu_0 - \lambda)^+\\
N&= \min\{n\geq 1: W_n > A  \}
\end{split}
\end{equation}
to detect the change. Here, $\lambda>0$ is the minimum amount of change in the mean of $h(X_{n-d}, \cdots, X_n)$ that the algorithm can detect, and is a design parameter. Before the change, the mean of the increment to $W_n$ is $\Expect[h(X_{n-d}, \cdots, X_n) - \mu_0 - \lambda] = - \lambda < 0$. Thus, like the CUSUM algorithm, the statistic $W_n$ here also has a negative drift before the change. Further, if $\Expect[h(X_{n-d}, \cdots, X_n)] > \mu_0 + \lambda$ after the change, then the drift will be positive, and the change can be detected using this algorithm. In the following, we call this algorithm the Deviation-CUSUM algorithm. 

We now provide some examples of summary statistics that can be employed in practice. 
\begin{enumerate}
\item \textit{Change in observation mean}: $h(X_{n-d}, \cdots, X_n) = X_n$. 
\item \textit{Change in variance}: Suppose the process is zero mean and $h(X_{n-d}, \cdots, X_n) = X_n^2$.
\item \textit{Change in entropy}: $h(X_{n-d}, \cdots, X_n) = \log 1/f_0(X_n)$, where $f_0$ is the pre-change density. 
\item \textit{Change in power spectrum}: Let $\{X_n\}$ be a zero mean stationary time-series. 
Then it has the spectral representation in terms of a Levy process $Z$ \cite{brockwell2013time}
\begin{equation}
X_n  = \int_{-\pi}^\pi e^{i n \nu} dZ(\nu).
\end{equation}
This implies that the autocorrelation function $\{R(h)\}$ of the process $X$ has the spectral representation
\begin{equation}
R(h) = \int_{-\pi}^\pi e^{i h \nu} dF(\nu),
\end{equation}
where $F$ is the quadratic variation of the process $Z$:  
\begin{equation}
dF = |dZ|^2.
\end{equation} 
The function $F$ is also called the power spectral density of the process $X$. In \cite{priestley1965evolutionary}, Priestley proposed a class of nonstationary processes with the representation
\begin{equation}
X_n  = \int_{-\pi}^\pi A_n(\nu) e^{i n \nu} dZ(\nu).
\end{equation}
Priestley called these processes locally stationary. The properties of the function $A_n(\nu)$ can be found in \cite{priestley1965evolutionary}. The quantity 
\begin{equation}
dF_n = |A_n|^2 dF
\end{equation}
is called the evolutionary spectrum of the process.
Then we can define
\begin{equation}\label{eq:spectrumH}
h(X_{n-d}, \cdots, X_n) = \int_\nu \hat{dF}_n(\nu),
\end{equation}
where $\hat{dF}_n$ is an estimate of the evolutionary spectrum using the variables $(X_{n-d}, \cdots, X_n)$ with mean
\begin{equation}
\mu_0 = \Expect[h(X_{n-d}, \cdots, X_n)] = \int_\nu dF(\nu)
\end{equation}
before change. There are other ways the summary statistic can be defined to capture a change in the power spectrum. We do not discuss them here. 

\end{enumerate}

\section{Behavioral Learning in Mice}\label{sec:MiceData}
\begin{figure}
\centering
\includegraphics[width=0.5\textwidth]{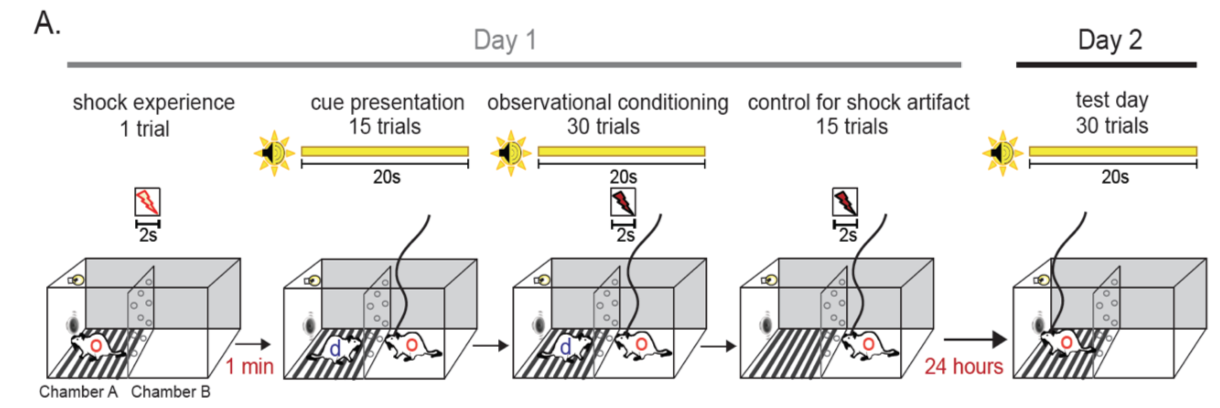}
\caption{Mice Experiment.}
\label{fig:MiceExpt}
\end{figure}

In a mice experiment, an observer mouse learns to associate a cue with a shock given to a demonstrator mouse. See Fig.\ref{fig:MiceExpt}. 
The details of the two-day experiment can be found in 
\cite{zhang2018estimating}. There were a total of 45 trials on Day 1. In the first 15 trials, a cue is not followed by shock. The firing pattern in these 15 trials is used as a baseline. In the next 30 trials, the cue is followed by a shock to the demonstrator mouse. Invasive data is collected from the observer mouse during the trial. The objective is to detect a possible change in neural firing pattern in the observer mouse after the shocks start in the trial 15, after the cue but before the shock is actually applied. This change in firing pattern from the baseline is seen as an indication of behavioral learning.  

\begin{figure}
\centering
\includegraphics[width=\linewidth]{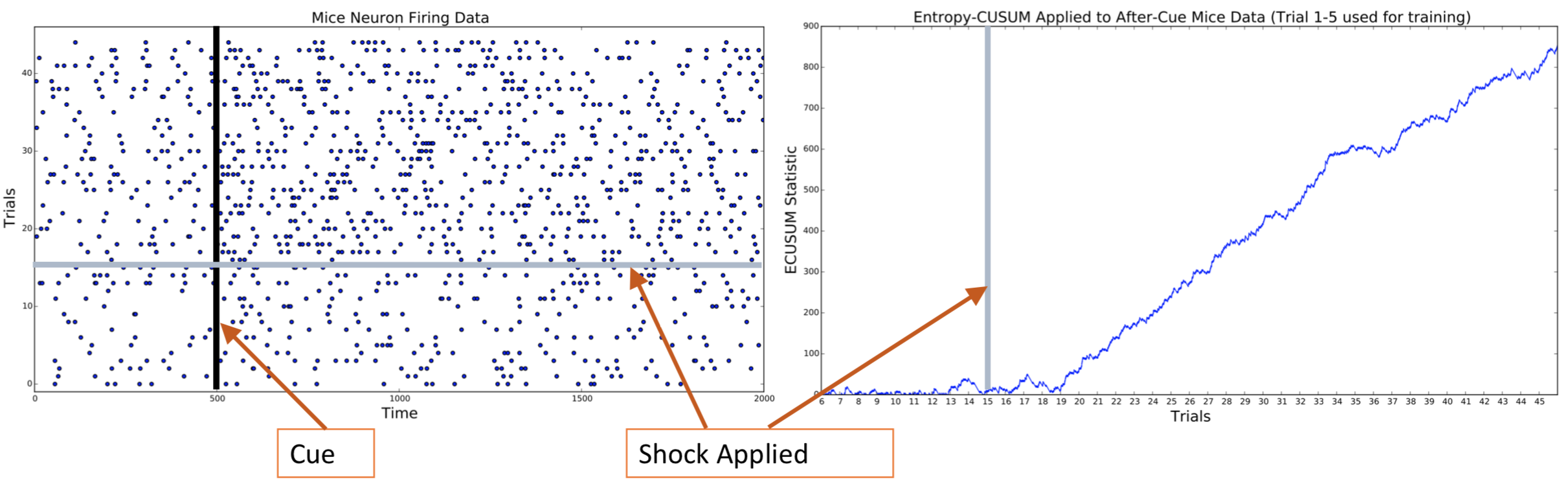}
\caption{The Deviation-CUSUM algorithm applied to binned spike data. The algorithm detects the change in the firing pattern.}
\label{fig:ECUSUM}
\end{figure}

In Fig.~\ref{fig:ECUSUM}, we have plotted the result of applying the Deviation-CUSUM algorithm to the mice data, also shown in the figure. The binner spike data is modeled 
as a Bernoulli process. The baseline is learned from the data from the first five trials. The sequence $\{X_n\}$ for the algorithm is obtained by concatenating the data from different trials as a single binary sequence. As seen in the figure, the algorithm successfully detect the change indicated by a change in the drift of the statistic. 

\begin{figure}
\centering
\includegraphics[width=\linewidth]{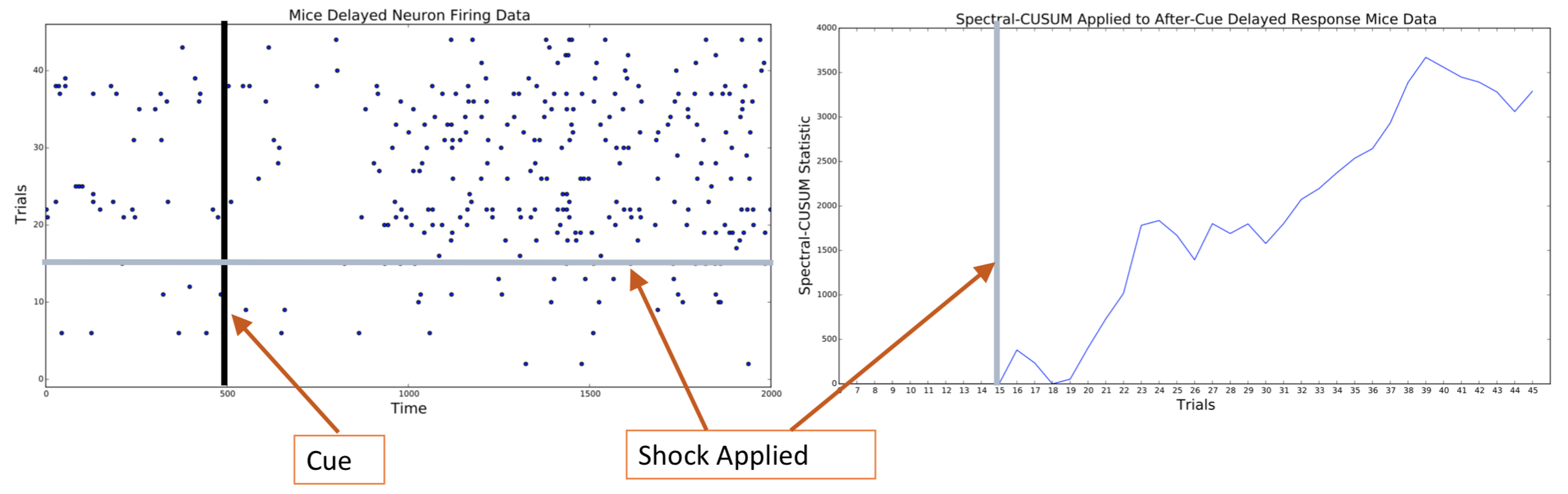}
\caption{The Deviation-CUSUM algorithm applied to delayed binned spike data. The algorithm detects the change in the firing pattern starting trial 15.}
\label{fig:SpectralCUSUM}
\end{figure}

In Fig.~\ref{fig:SpectralCUSUM}, we have shown spike data where the response is delayed. For this type of data, it may be hard to detect the change using the techniques used for Fig.~\ref{fig:ECUSUM} due to a lack of persistent firing. As a result, we apply the spectrum based technique \eqref{eq:spectrumH} to the data. Here, to obtain the spectrum, we used $d$ equal to the length of a trial. Also, the value of $W_n$ is computed only at the beginning of the trials. Hence, the time index here is trials rather than the bin-level slots. The baseline is again learned from the first five trials. As seen in the figure, the Deviation-CUSUM detects the change in firing pattern here as well. 

\bibliographystyle{ieeetr}

\bibliography{QCD_verSubmitted}

\balance

\end{document}